\documentclass[11pt]{article}
\usepackage{hyperref}
\usepackage{amssymb}
\usepackage{graphicx}
\usepackage{amsmath}
\usepackage{authblk}
\usepackage{cite}
\usepackage{physics}
\usepackage{revsymb}
\usepackage{float}
\usepackage{slashed}
\usepackage{tabularx}
\usepackage{breqn}
\usepackage{array}
\usepackage{verbatim}
\usepackage{dsfont}
\usepackage[section]{placeins}
\usepackage[a4paper, total={6.8in, 10in}]{geometry}
\numberwithin{equation}{section}

\begin{document}
	\title{Operator formulation of Classical mechanics: Levi-Civita map and equivalence of central forces in 2-dimensions}
	\author{ {\bf {\normalsize Nikhil Kasyap Puranam}\thanks{nikhilkasyap547@gmail.com},\,
			{\bf {\normalsize E. Harikumar}\thanks{eharikumar@uohyd.ac.in}}}\\
		{\normalsize School of Physics, University of Hyderabad}\\{\normalsize Central University P.O, Hyderabad-500046, Telangana, India}\\ }
	\date{}

	\maketitle
	\begin{abstract}

	We study the operator formulation of classical mechanics by explicitly applying it to two central potentials in 2 dimensions. After constructing the classical Hamiltonian operators and corresponding Schr\"odinger like equations, we solve for the corresponding classical wave functions associated with these two potentials, viz; Kepler and harmonic potentials. While satisfying continuity equations, these classical wave functions are shown to be renormalizable only in a finite region of the 2D plane. We also derive the well-known equivalence between these two models within the operator formulation of classical mechanics. This equivalence is shown by relating the Schr\"odinger-like equations and corresponding classical wave functions of these two systems, using the Levi-Civita map and a reparametrizaton of time(Sundman map).

	\end{abstract}
	%\newpage
	\section{Introduction}

	The essential mathematical structure describing quantum phenomena are vector spaces and operators with well-defined action on these vector spaces. This framework is alien to the conventional formalism of classical mechanics and attempts to reformulate classical physics in terms of operators and corresponding vector spaces has a long history\cite{jc,bk,vn}.
	Among the approaches developed in this direction, the most popular one is the Koopman-von Neumann(KvN) framework \cite{dm}. In this approach, the wave function describing a classical system is defined in the corresponding phase space, unlike in quantum mechanics, where it is defined in the configuration space. Further, the position and momentum operators commute in this approach, as expected for a classical system. Thus, the parallel between KvN mechanics and quantum mechanics starts and ends with the use of vector spaces alone.
	
	In another approach \cite{jc}, the classical wave function is entirely defined in the configuration space, and the position and momentum operators do not commute. The Hamiltonian operator introduced in this formulation differs from the classical Hamiltonian, but the corresponding Schr\"{o}dinger-like equation describing the time evolution of the classical wave function reproduces the usual Hamilton-Jacobi equation and a particle number conservation equation. Thus, this approach shares specific characteristics of classical and quantum mechanics.

	In this approach\cite{jc}, one starts with the Hamilton-Jacobi equation
	\begin{equation}\label{HJeqn}
		\displaystyle \dfrac{1}{2m} (\nabla S)^{2} + V =E \; ,
	\end{equation}
where $E=\dfrac{\partial S}{\partial t}$ and define classical wave function in terms of Hamilton's principle function $S$ as
	\begin{equation} \label{wf}
		\displaystyle \psi ({\vec r},t) = \rho ({\vec r},t) e^{\frac{i}{\hbar}S({\vec r},t)}.
	\end{equation}
This classical wave function describes the time evolution of a group of identical, non-interacting particles whose dynamics are governed by the Hamilton-Jacobi equation. Note that $\psi^*\psi=\rho^2$ gives the probability density of this group of non-interacting particles in the configuration space. The normalisation condition that this probability density satisfies is $\int |\psi|^{2} dx = \int \rho ^2 dx =1$. The conservation of particle number (in the group of particles) in the configuration space is prescribed by the continuity equation satisfied by $\rho ({\vec r},t)$ and $S({\vec r},t)$ and is given by
	\begin{equation} \label{cont}
		\displaystyle \nabla \cdot (\rho ^{2}\nabla S) = -m \dfrac{\partial \rho ^{2}}{\partial t} \;.
	\end{equation}
In this approach, one defines a classical Hamiltonian operator
\begin{equation} \label{Hclass}
	\displaystyle	\hat{\mathcal{H}}_{cl} = -\dfrac{\hbar ^{2}}{2m}\nabla ^{2} + V + \dfrac{\hbar ^{2}}{2m}\dfrac{\nabla ^{2}\rho}{\rho} = \hat{\mathcal{H}}+\frac{1}{2m\rho}\nabla^2\rho ,
	\end{equation}
one immediately verifies that the real and imaginary part  of the Schr\"odinger like equation
 \begin{equation}\label{SLE}
	 \hat{\mathcal{H}}_{cl}\psi = i\hbar \dfrac{\partial \psi}{\partial t}\;
	 \end{equation}
 are the same as the Hamilton-Jacobi equation and the continuity equation. Thus, the classical Hamiltonian operator in eqn.(\ref{Hclass}), classical wave function in eqn.(\ref{wf}), along with the Schr\"{o}dinger-like equation in eqn.(\ref{SLE}) provides an equivalent description of the classical system satisfying the Hamilton-Jacobi equation in eqn.(\ref{HJeqn}). Note that the $\hbar$ appearing in eqn(\ref{Hclass}) will cancel with $\hbar$ coming from $\psi$ and thus eqn.(\ref{HJeqn}) and eqn.(\ref{cont}) will be independent of $\hbar$. From now onwards, we set $\hbar = 1$.

 The operator $\frac{1}{2m\rho}\nabla^2\rho$ appearing in the classical Hamiltonian operator(see eqn.(\ref{Hclass})) makes it different from the Hamiltonian operator of usual quantum mechanics. It is important to note that this operator is non-linear and, in general, non-hermitian. In this operator framework of classical mechanics, the phase space coordinates $q_i$ and $p_i$ do not commute, and they also satisfy an uncertainty relation as in quantum mechanics. Status of Ehrenfest theorem has also been analysed in\cite{jc}.

	In this paper, we first explicitly construct the Schr\"odinger-like equation obeyed by the classical wave function associated with the classical Kepler problem in 2 dimensions and the continuity equation satisfied by the probability density. We then solve the  Schr\"odinger like equation and obtain the
	classical wave function. Both these solutions are derived in the polar coordinates. We then discuss the normalisation of the classical wave function. We show that the normalisation condition satisfied here is different from the corresponding quantum mechanical wave function in a crucial aspect: the range over which one integrates $\psi^2\psi$ in the present case is finite, unlike in the case of the quantum mechanical counterpart. We then repeat this study for a 2-dimensional harmonic oscillator.

	After setting up the Kepler problem and harmonic oscillator in 2-dimensions in the operator formalism of classical mechanics, we then study the status of the well-known equivalence between these two systems\cite{lc}. We re-derive this celebrated equivalence within the operator formalism here. We show that this equivalence does exist even in this operator framework, irrespective of the non-linear and non-hermitian nature of $\hat{\mathcal{H}}_{cl}$. We emphasize that the result reported here is that the equivalence between these two central force systems in 2d is re-derived here, using the formulation of these problems as classical operator problems. We show that the classical Schr\"{o}dinger equation describing these two problems is equivalent under the Levi-Civita map. Thus, our result validates the operator formalism.
	
	This paper is organized as follows. In the next section, we summarize the operator formulation of the Kepler problem in 2 dimensions. Then we explicitly write down the Schr\"{o}dinger like equation and obtain the corresponding classical wave function associated with the Kepler problem. Section 3 presents the operator formulation of the 2d harmonic oscillator. In both these sections, we use plane polar coordinates. Our main results are presented in Section 4. In this section, we apply the Levi-Civita map adapted to the plane-polar coordinates and the reparameterisation of time to show that the Hamilton-Jacobi equation and continuity equation corresponding to the 2d Kepler problem are mapped to those of a 2d harmonic oscillator. We then show that under certain conditions, the classical wave functions of two systems get mapped. Since $\hat{\mathcal{H}}_{cl}\psi=i\dfrac{\partial \psi}{\partial t}$ leads to Hamilton-Jacobi equation and continuity/conservation equation on substitution of the wave function, the Schr\"{o}dinger like equation corresponding to the Kepler problem gets mapped to that of harmonic oscillator. We conclude the section by explicitly showing this equivalence of the Schr\"{o}dinger like equation for both problems on constant energy surfaces. We present our concluding remarks in Section 5.
	
	\section{Operator formulation of the 2d Kepler Problem}

	We start with the Hamiltonian for the Kepler problem written in plane polar coordinates
	\begin{equation}
		\displaystyle \mathcal{H} = \dfrac{p_{r}^{2}}{2m} + \dfrac{p_{\phi}^{2}}{2mr^{2}} - \dfrac{k}{r} \; . \label{1}
	\end{equation}
	Here $p_{r} = m \frac{d r}{d t}$ and $p_{\phi}= m r^{2}\frac{d \phi}{dt} = l$ is the conserved angular momentum. The corresponding Hamilton-Jacobi equation is
	\begin{equation}
		\displaystyle  \dfrac{1}{2m}\left(\dfrac{\partial S}{\partial r}\right)^{2} + \dfrac{1}{2mr^{2}}\left(\dfrac{\partial S}{\partial \phi}\right)^{2} - \dfrac{k}{r} +\dfrac{\partial S}{\partial t} = 0 \;, \label{5}
	\end{equation}
	where $p_{r}= \dfrac{\partial S}{\partial r}$ and $p_{\phi}=\dfrac{\partial S}{\partial \phi}$. Substituting for Hamilton's principle function
	\begin{equation}\label{hpf}
	\displaystyle S = R(r)+l\phi-Et
    \end{equation}
    In the above equation, solving for $R(r)$ gives
    \begin{equation}\label{7}
	\displaystyle R(r) = \int dr \sqrt{2mE+\dfrac{2mk}{r}-\dfrac{l^{2}}{r^{2}}}.
\end{equation}
    Using this back in $ R(r)=S-l\phi-Et$  and noting $t=\frac {\partial R(r)}{\partial t}$ we find
	\begin{equation}
		\displaystyle t = \int m dr \left(\dfrac{d R}{d r}\right)^{-1}_{.}\label{11}
		\end{equation}
Since $\phi=\frac {\partial R(r)}{\partial l}$ we also get
\begin{equation}
 \phi = \int \dfrac{l}{r^{2}} dr \left(\dfrac{d R}{d r}\right)^{-1}_{.} 	\label{112}
\end{equation}
The continuity equation(see eqn.(\ref{cont}) $\displaystyle \nabla \rho \cdot \nabla S +\dfrac{\rho}{2}\nabla ^{2}S = -m \dfrac{\partial \rho}{\partial t}$ in plane-polar coordinates is,
	\begin{equation}\label{3.6}
		\displaystyle \dfrac{\partial S}{\partial r}\dfrac{\partial \rho}{\partial r}+\dfrac{1}{r^{2}}\dfrac{\partial \rho}{\partial \phi}\dfrac{\partial S}{\partial \phi}+\dfrac{\rho}{2}\left(\dfrac{1}{r}\dfrac{\partial}{\partial r}\left(r\dfrac{\partial S}{\partial r}\right)+\dfrac{1}{r^{2}}\dfrac{\partial ^{2}S}{\partial \phi ^{2}}\right) = -m\dfrac{\partial \rho}{\partial t}.
	\end{equation}
	Using the method of separation of variables, we find $\rho$ to be
	\begin{align}\label{prob}
		\rho (r,\phi , t) &= X(r)Y(\phi)T(t) \nonumber \\
		&=  \dfrac{C}{\sqrt{r\frac{\partial S}{\partial r}\frac{\partial S}{\partial \phi}}}\;\; exp\left(-\frac{\alpha}{m}\left(t-\int dr \dfrac{1}{\frac{\partial S}{\partial r}}\right)-\beta\left(\int dr \dfrac{1}{r^{2}\frac{\partial S}{\partial r}}-\int \frac{1}{\frac{\partial S}{\partial \phi}}d\phi\right)\right),
	\end{align}
	where $C,\alpha,\beta$ are constants. Using eqn.(\ref{11})and eqn.(\ref{112}) in the exponent on the RHS of the above equation, we find,
	\begin{equation}\label{3.7}
		\displaystyle \rho (r,\phi , t) =\dfrac{C}{\sqrt{r\frac{\partial S}{\partial r}\frac{\partial S}{\partial \phi}}}
		= \dfrac{C}{\sqrt{rl}\left(2mE+\dfrac{2mk}{r}-\dfrac{l^{2}}{r^{2}}\right)^{1/4}},
	\end{equation}
	where we have used $\dfrac{\partial S}{\partial r}=\dfrac{dR}{dr}$ and $\dfrac{\partial S}{\partial \phi}=l$. Using the above $\rho$ in eqn.(\ref{Hclass}), we find the Schrodinger like equation(see eqn.(\ref{SLE}) for Kepler problem to be
	\begin{equation}
		\left(-\dfrac{1}{2m}\nabla ^{ 2}-\dfrac{k}{r}+\dfrac{1}{2m}\dfrac{\nabla ^{ 2}\rho }{\rho }\right)\psi = i \dfrac{\partial \psi}{\partial t},
	\end{equation}
	The explicit form of the wave function (see eqn.(\ref{wf})) satisfying the above equation is
	\begin{equation}\label{cwf}
		\psi (r,\phi,t) =   \dfrac{C}{\sqrt{rl}\left(2mE+\dfrac{2mk}{r}-\dfrac{l^{2}}{r^{2}}\right)^{1/4}}\;\;\; e^{i(R(r)+l\phi-Et)} \;\; .
	\end{equation}
	 Note that  $\rho$ and $S$ obtained from eqn.(\ref{3.7}) and eqn.(\ref{cwf}), respectively, satisfy continuity equation(see eqn.\ref{cont}).
	
	Since we are dealing with Classical systems, when normalizing the classical wave functions, we should only consider the region that is classically accessible to the particle. For the Kepler potential at turning points, the radial velocity and equivalently, the kinetic energy become zero, and the total energy is equal to the effective potential energy. Thus $E = -\dfrac{k}{r}+\dfrac{l^2}{2mr^2}$, which after rearranging becomes
	\begin{equation}
	 2mE+\dfrac{2mk}{r}-\dfrac{l^2}{r^2}=0 .
	\end{equation}
Thus, we find the turning points for the Kepler problem are
	\begin{equation}\label{rpm}
	 r_{\pm} = \dfrac{-k}{2E} \pm\dfrac{k}{2E}e,
	\end{equation}
	where $e= 1+\frac{2l^2E}{mk^2}$ is the eccentricity of the orbit. Also, notice that the classical wave function is only normalizable for bound states, that is, for elliptical and circular orbits. Therefore, for elliptical and circular orbits, the above equation gives limits of integration for normalization as $r_{\pm}$ and thus we have $\int_{r_-}^{r^+}\psi^*\psi r drd\phi=1$. We note that the
	quantum mechanical wavefunction corresponding to the 2d hydrogen atom \cite{2h}
	\begin{equation}
	 \Psi (r,\phi ,t)_{QM} = A (\beta _n r)^{|l|}G(\beta _n r)e^{i(l\phi - Et)}
	\end{equation}
	where $\beta _n = \dfrac{n}{2mZe^2}$, $G(\beta _n r)$ is the confluent hypergeometric function, and $A$ is the normalisation constant. Notice that the difference between the classical wave function (\ref{cwf}) and the above is in the radial part and the normalisation factor.

	\section{Operator formulation of the 2d Harmonic Oscillator}

	The Hamiltonian for the 2d harmonic oscillator in plane polar coordinates is
\begin{equation}\label{13}
	\displaystyle \mathcal{H}^{\prime} = \dfrac{p_{\tilde{r}}^{2}}{2m} + \dfrac{p_{\theta}^{2}}{2m\tilde{r}^{2}} + \dfrac{1}{2}k^{\prime}\tilde{r}^{2}.
\end{equation}
Here,
$	p_{\tilde{r}} = m \frac{d \tilde{r}}{d\tau},	p_{\theta}=  m\tilde{r}^{2}\frac{d \theta}{d \tau}= l^{\prime}$ is the conserved angular momentum. Note that the time parameter appearing here differs from the time parameter used in the description of the 2d Kepler problem. The Hamilton-Jacobi equation corresponding to eqn.(\ref{13}),
 $\mathcal{H}^{\prime}\left(q^{\prime}_{i},\dfrac{\partial \tilde{S}}{\partial q^{\prime}_{i}},\tau\right) + \dfrac{\partial \tilde{S}}{\partial \tau} = 0 $ is
\begin{equation}
	\displaystyle  \dfrac{1}{2m}\left(\dfrac{\partial \tilde{S}}{\partial \tilde{r}}\right)^{2} + \dfrac{1}{2m\tilde{r}^{2}}\left(\dfrac{\partial \tilde{S}}{\partial \theta}\right)^{2} + \dfrac{1}{2}k^{\prime}\tilde{r}^{2} +\dfrac{\partial \tilde{S}}{\partial \tau} = 0 \label{17}
\end{equation}

Here $p_{\tilde{r}} =\dfrac{\partial \tilde{S}}{\partial \tilde{r}} \; , p_{\theta} = \dfrac{\partial \tilde{S}}{\partial \theta} \label{15}$ and $\tilde{S}$ is the Hamilton's principal function. Using,
\begin{equation}\label{hpf2}
	\displaystyle \tilde{S} = \tilde{R}(\tilde{r})+l^{\prime}\theta-E^{\prime}\tau,
\end{equation}
repeating the same steps in deriving eqns.(\ref{7}, \ref{11}, \ref{112}), we obtain for 2-dim harmonic oscillator
\begin{align}\label{23}
	\displaystyle \tau &= \int m d\tilde{r} \left(\dfrac{d \tilde{R}}{d \tilde{r}}\right)^{-1}_, \\
 \theta &= \int \dfrac{l^{\prime}}{\tilde{r}^{2}} d\tilde{r} \left(\dfrac{d \tilde{R}}{d \tilde{r}}\right)^{-1}_.
\end{align}
Solving the corresponding continuity equation (see eqn.(\ref{cont})) in plane polar coordinates gives
\begin{equation}\label{1160}
	\tilde{\rho} (\tilde{r},\theta , \tau) =\dfrac{C^{\prime}}{\sqrt{\tilde{r}l^{\prime}p_{\tilde{r}}}}=\dfrac{C^{\prime}}{\sqrt{\tilde{r}l^{\prime}}\left(2mE^{\prime}-mk^{\prime}\tilde{r}^{2}-\dfrac{l^{\prime 2}}{\tilde{r}^{2}}\right)^{1/4}}.
\end{equation} 
The classical wave function for the harmonic oscillator in 2 dimensions $
\Phi ({\vec r},\tau) = {\tilde\rho} ({\vec r},\tau ) e^{\frac{i}{\hbar}{\tilde {S}({\vec r},\tau)}}$ satisfy the Schr\"{o}dinger like equation \cite{jc} for simple harmonic oscillator given by
\begin{equation}
	\left(-\dfrac{1}{2m}\nabla^{\prime 2}+\dfrac{1}{2}k^{\prime}\tilde{r}^{2}+\dfrac{\nabla^{\prime 2}\tilde{\rho}}{\tilde{\rho}}\right)\Phi = i \dfrac{d\Phi}{d\tau},
\end{equation}
where $\nabla^{\prime 2}$ is the Laplacian in $({\tilde r}, \theta)$ coordinates.
The classical wave function is explicitly given by
\begin{equation}\label{4.7}
	\Phi(\tilde{r},\theta ,\tau) = \dfrac{C^{\prime}}{\sqrt{\tilde{r}l^{\prime}}\left(2mE^{\prime}-mk^{\prime}\tilde{r}^{2}-\dfrac{l^{\prime 2}}{\tilde{r}^{2}}\right)^{1/4}} \;\;\; e^{i(\tilde{R}(\tilde{r})+l^{\prime}\theta-E^{\prime}\tau)}.
\end{equation} 
As for the Kepler problem, here too we normalize the wave function, taking into account the classical accessible region. The turning points of Harmonic motion bound this finite region.
From the condition satisfied at the turning points (see discussion after eqn.(\ref{cwf}))
\begin{equation}
 2mE^{\prime}-mk^{\prime}\tilde{r}^{2}-\dfrac{l^{\prime 2}}{\tilde{r}^{2}}=0\;\;,
\end{equation}
we find
\begin{equation}\label{rtildepm}
 \tilde{r}_{\pm} = \left(\dfrac{E^{\prime}}{k^{\prime}}\pm \dfrac{E^{\prime}}{k^{\prime}}    \sqrt{1-\dfrac{l^{\prime 2}k^{\prime}}{m E^{\prime 2}}}\right)^{1/2}_.
\end{equation}
We have neglected the negative root in the above because the radial coordinate can only take positive values. Therefore, the above positive roots will set the limits of integration for normalization of $\Phi$, i.e., the normalisation condition is $\int_{{\tilde r}_{-}}^{{\tilde r}_{+}} \Phi^*\Phi \tilde r d{\tilde r}d\theta=1$.  Here to the
 quantum mechanical result of 2d harmonic oscillator in polar coordinates\cite{bj}
\begin{equation}
 \Upsilon (\tilde{r},\theta ,\tau)= B\;\; (\alpha \tilde{r})^{|l^{\prime}|}e^{(\alpha \tilde{r})^{2}/2}g(\alpha \tilde{r})e^{i(l^{\prime}\theta - E^{\prime}\tau)},
\end{equation}
where $\alpha = (m\omega)^{1/4}$, $g(\alpha \tilde{r})$ is the confluent hypergeometric function, and $B$ is the normalisation constant. Notice that the above equation differs from the classical wave function (\ref{4.7}) in the radial part and the normalisation factor.

\section{Levi-Civita map and Reparameterization of time in plane Polar Coordinates}

In this section, we use the Levi-Civita map and reparametrisation of time parameter $t$ and show that the Hamilton-Jacobi equation, continuity equation, probability density, classical wave function, and the Schr\"odinger-like equation corresponding to the Kepler problem in 2-dimensions are mapped to the corresponding one of a harmonic oscillator in 2-dimension. This establishes the well-known equivalence between these two systems, now within the framework of operator formalism. Thus, we show that the operator formalism, which has some features common with the framework of classical mechanics and some common with quantum mechanics, is robust in maintaining the well-known equivalence. This is important as the normalisation of classical wave functions and the classical Hamiltonian operators of these two systems differ from the corresponding quantities in the quantum mechanical treatment.

2d Kepler problem and 2d harmonic oscillator are equivalent as the equation of motion of the former is mapped to that of the latter by the Levi-Civita map and time reparametrisation\cite{lc,eh1}. Equation of motion describing the Kepler problem in 2d, written in terms of complex coordinates $Z$, is mapped to that of a 2d harmonic oscillator written in terms of complex coordinates $U$ by the transformation given by

	\begin{equation}\label{1156}
	\displaystyle	Z = \gamma U^{2}
	\end{equation}
	along with a reparameterization of the time variable given by
	\begin{equation}\label{1139}
		dt = \dfrac{r}{c}d\tau = \dfrac{\tilde{r}^{2}}{c}d\tau.
	\end{equation}
	Here, $c$ is a constant which we will fix later. We write both $Z$ and $U$ in polar form 
	\begin{equation}
		Z = r e^{i\phi},~~
		U = \tilde{r}e^{i\theta}.
	\end{equation}
	Since the constant $\gamma$ in eqn.(ref{1156}) is introduced for dimensional consistency, it can be set equal to $1$ without loss of generality. Substituting $Z,U$ in eqn.(\ref{1156}), we get the following identifications
	\begin{align}
		r &= \tilde{r}^{2},  \label{1141} \\
		\phi &= 2\theta. \label{1142}
	\end{align}
	
We now use the Levi-Civita map, the time reparameterization, re-expressing radial and angular momentum, and the energy of the 2d Kepler problem in terms of the new coordinates and time parameter.

Radial momentum for the Kepler Problem becomes
\begin{equation}\label{1135}
	p_{r} = m \dfrac{dr}{dt} 
	 = m \dfrac{c}{\tilde{r}^{2}}\dfrac{d\tilde{r}^{2}}{d\tau} = \dfrac{2c}{\tilde{r}} p_{\tilde{r}},
\end{equation}
where $p_{\tilde{r}} = m \dfrac{d\tilde{r}}{d\tau}$.
Similarly, we map the angular momentum for the Kepler problem
\begin{equation}\label{1136}
	p_{\phi} = mr^{2}\dfrac{d\phi}{dt} = m \tilde{r}^{4}\dfrac{c}{\tilde{r}^{2}}\dfrac{d(2\theta)}{d\tau} = 2c p_{\theta},
\end{equation}
where $ p_{\theta} = m \tilde{r}^{2}\dfrac{d\theta}{d\tau}$. The energy for the Kepler problem
\begin{equation}
	\displaystyle E = \dfrac{1}{2m}\left(p_{r}^{2}+\dfrac{p_{\phi}^{2}}{2mr^{2}}\right)-\dfrac{k}{r},
\end{equation}
is now re-expressed using eqn.(\ref{1135}) and eqn.(\ref{1136}) as
\begin{equation} \label{1137}
	E =  \left(\dfrac{2c}{\tilde{r}}\right)^{2}\left[\dfrac{1}{2m}\left(p_{\tilde{r}}^{2}+\dfrac{p_{\theta}^{2}}{\tilde{r}^{2}}\right)\right]-\dfrac{k}{\tilde{r}^{2}}.
\end{equation}
In terms of the total energy of the 2d Harmonic Oscillator 
\begin{equation}
	\displaystyle E^{\prime} = \dfrac{1}{2m}\left(p_{\tilde{r}}^{2}+\dfrac{p_{\theta}^{2}}{\tilde{r}^{2}}\right)+\dfrac{1}{2}k^{\prime}\tilde{r}^{2} \; ,
\end{equation}
we re-express eqn.(\ref{1137}) as
\begin{equation}
	E = \left(\dfrac{2c}{\tilde{r}}\right)^{2}\left[E^{\prime}-\dfrac{1}{2}k^{\prime}\tilde{r}^{2}\right]-\dfrac{k}{\tilde{r}^{2}} \; ,
\end{equation}
which on re-arranging becomes
\begin{equation}\label{5.13}
 	E+\dfrac{k}{\tilde{r}^{2}} = \dfrac{4c^{2}E^{\prime}}{\tilde{r}^{2}}-2c^{2}k^{\prime}.
\end{equation}
Comparing the coefficients of terms with the same power of $\tilde{r}$ on both sides of the above equation, we obtain the following identifications 
\begin{equation}\label{500}
	\displaystyle E = -2c^{2}k^{\prime}; \;\;\;  k = 4c^{2}E^{\prime}.
\end{equation}
To map the Hamilton-Jacobi equation describing 2d Kepler problem to that of 2d harmonic oscillator we need to re-express the time derivative of the action corresponding to 2d Kepler problem in terms of the time derivative of the action of 2d harmonic oscillator. For this we start with
\begin{align}
	dS &= dR(r) + ld\phi -E dt, \label{5.15}
\end{align}
 where $S$ is Hamilton's principal function for the Kepler problem, which satisfies (see eqn.(\ref{hpf})). Using eqn.(\ref{500}) (see eqn.(\ref{7})), the first term on the RHS of the above equation becomes
\begin{align}
	dR &= dr \sqrt{2mE+\dfrac{2mk}{r}-\dfrac{l^{2}}{r^{r}}} 
	= 2\tilde{r} d\tilde{r} \sqrt{2m(-2c^{2}k^{\prime})+\dfrac{2m(4c^{2}E^{\prime})}{\tilde{r}^{2}}-\dfrac{4c^{2}l^{\prime 2}}{\tilde{r}^{4}}} \nonumber \\
	&= 2\tilde{r}d\tilde{r}\sqrt{\dfrac{4c^{2}}{\tilde{r}^{2}}}\sqrt{2mE^{\prime}-mk^{\prime}\tilde{r}^{2}-\dfrac{l^{\prime}}{\tilde{r}^{2}} }
	= 4c\;\; d\tilde{r} \sqrt{2mE^{\prime}-mk^{\prime}\tilde{r}^{2}-\dfrac{l^{\prime 2}}{\tilde{r}^{2}} }
	 = 4c\; d\tilde{R}(\tilde{r}), \label{5.16}
\end{align}
In the above, we have also used the identification (see eqn.(\ref{1136}))
\begin{equation}
	ld\phi = 4cl^{\prime}d\theta. \label{5.17}
\end{equation}
 With these we find,
\begin{equation}
	dS 	= 4c(d\tilde{R}+l^{\prime}d\theta)+2c^{2}k^{\prime}\dfrac{\tilde{r}^{2}}{c}d\tau 
	 = 4c\left(d\tilde{S}+\left(E^{\prime}+\dfrac{k^{\prime}\tilde{r}^{2}}{2}\right)d\tau\right),
\end{equation} 

where $\tilde{S}$ is the Hamilton's principal function corresponding to 2d harmonic oscillator. Further, $E^{\prime},k^{\prime},\tilde{r},\tau$ are all quantities associated with 2d harmonic oscillator. Differentiating both sides with respect to $\tau$, we get
\begin{equation}\label{28}
	\dfrac{\partial S}{\partial \tau} = 4c\left(\dfrac{\partial \tilde{S}}{\partial \tau}+\left(E^{\prime}+\dfrac{k^{\prime}\tilde{r}^{2}}{2}\right)\right),
\end{equation}
which gives the relation between the time derivatives of action of 2d Kepler and 2d harmonic oscillator.

\subsection{Transformation of Hamilton-Jacobi equations}

We now show that under the mapping between $p_{r},p_{\phi}, E,k$ and $\ frac {\partial S}{\partial \tau}$, the Hamilton-Jacobi equation describing the 2d Kepler problem gets mapped to that of a 2d harmonic oscillator. Using eqn.(\ref{1135}) and eqn.(\ref{1136}), we see that
\begin{equation}
	p_{r} = \dfrac{\partial S}{\partial r} = \dfrac{2c}{\tilde{r}}\dfrac{\partial \tilde{S}}{\partial \tilde{r}} = \dfrac{2c}{\tilde{r}}p_{\tilde{r}} \; ,
\end{equation}
\begin{equation}
 p_{\phi} =	\dfrac{\partial S}{\partial \phi}= 2c\dfrac{\partial \tilde{S}}{\partial \theta} = 2cp_{\theta} \; .
\end{equation}
Using these in eqn.(\ref{5}), we find
\begin{equation}
	\dfrac{1}{2m}\left(\dfrac{2c}{\tilde{r}}\right)^{2}\left[\left(\dfrac{\partial \tilde{S}}{\partial \tilde{r}}\right)^{2}+\dfrac{1}{\tilde{r}^{2}}\left(\dfrac{\partial \tilde{S}}{\partial \theta}\right)^{2}\right]-\dfrac{k}{\tilde{r}^{2}}+\dfrac{\partial S}{\partial t} = 0\; .
\end{equation}
Applying time reparameterization(see eqn.(\ref{1139})) to the last term in the above equation, we get,
\begin{equation}
		\dfrac{1}{2m}\left(\dfrac{2c}{\tilde{r}}\right)^{2}\left[\left(\dfrac{\partial \tilde{S}}{\partial \tilde{r}}\right)^{2}+\dfrac{1}{\tilde{r}^{2}}\left(\dfrac{\partial \tilde{S}}{\partial \theta}\right)^{2}\right]-\dfrac{k}{\tilde{r}^{2}}+\dfrac{c}{\tilde{r}^{2}}\dfrac{\partial S}{\partial \tau} = 0 \; .
\end{equation}
Using eqn.(\ref{28}) this equation becomes,
\begin{equation}
		\dfrac{1}{2m}\left(\dfrac{2c}{\tilde{r}}\right)^{2}\left[\left(\dfrac{\partial \tilde{S}}{\partial \tilde{r}}\right)^{2}+\dfrac{1}{\tilde{r}^{2}}\left(\dfrac{\partial \tilde{S}}{\partial \theta}\right)^{2}\right]-\dfrac{k}{\tilde{r}^{2}}+\dfrac{4c^{2}}{\tilde{r}^{2}}\left(\dfrac{\partial \tilde{S}}{\partial \tau}+E^{\prime}+\dfrac{k^{\prime}\tilde{r}^{2}}{2}\right) = 0\; .
\end{equation}
Since $-\dfrac{k}{\tilde{r}^{2}}+\dfrac{4c^{2}}{\tilde{r}^{2}}E^{\prime}=0$, this reduces to
	\begin{equation}
			\displaystyle \dfrac{1}{2m}\left[\left(\dfrac{\partial \tilde{S}}{\partial \tilde{r}}\right)^{2}+\dfrac{1}{\tilde{r}^{2}}\left(\dfrac{\partial \tilde{S}}{\partial \theta}\right)^{2}\right]+\dfrac{k^{\prime}\tilde{r}^{2}}{2}+\dfrac{\partial \tilde{S}}{\partial \tau} = 0 \; ,
	\end{equation}
	which is the Hamilton-Jacobi equation for a simple harmonic oscillator given in eqn.(\ref{17})

	\subsection{Transformation of Continuity Equations}

Applying the mappings given in eqn.(\ref{1141}) and eqn.(\ref{1139}) to the first term of continuity equation
$\displaystyle \nabla \rho \cdot \nabla S +\dfrac{\rho}{2}\nabla ^{2}S = -m \dfrac{\partial \rho}{\partial t}$ we find
	\begin{equation}
		\nabla \rho \cdot \nabla S = \dfrac{\partial \rho}{\partial r}\dfrac{\partial S}{\partial r}+\dfrac{1}{r^{2}}\dfrac{\partial \rho}{\partial \phi}\dfrac{\partial S}{\partial \phi}  
        = \dfrac{c}{\tilde{r}^{2}}\left[\dfrac{\partial \rho}{\partial \tilde{r}}\dfrac{\partial \tilde{S}}{\partial \tilde{r}}+\dfrac{1}{\tilde{r}^{2}}\dfrac{\partial \rho}{\partial \theta}\dfrac{\partial \tilde{S}}{\partial \theta}\right].
	\end{equation}
	Similarly we find
	\begin{equation}
		\nabla ^{2} S = \dfrac{1}{r}\dfrac{\partial }{\partial r}\left(r\dfrac{\partial S}{\partial r}\right)+\dfrac{1}{r^{2}}\dfrac{\partial ^{2}S}{\partial \phi ^{2}} 
		 = \dfrac{c}{\tilde{r}^{2}}\left[\dfrac{1}{\tilde{r}}\dfrac{\partial}{\partial \tilde{r}}\left(\tilde{r}\dfrac{\partial \tilde{S}}{\partial \tilde{r}}\right)+\dfrac{1}{\tilde{r}^{2}}\dfrac{\partial ^{2}\tilde{S}}{\partial \theta ^{2}}\right].
	\end{equation}
	Combining both of these, we find that the LHS of the continuity equation becomes
	\begin{equation}
		\nabla \rho \cdot \nabla S +\frac{\rho}{2} \nabla ^{2}S = \dfrac{c}{\tilde{r}^{2}}\left[\nabla ^{\prime}\rho \cdot \nabla ^{\prime}\tilde{S}+\dfrac{\rho}{2}\nabla ^{\prime 2}\tilde{S}\right] \; .
	\end{equation}
	Here $\nabla ^{\prime}$ denotes differentiation with respect to variables used in the description of a simple harmonic oscillator $(\tilde{r},\theta)$. Now, the right side of the continuity equation, after reparameterization of time, gives
	\begin{equation}
		m\dfrac{\partial \rho}{\partial t} = m\dfrac{c}{\tilde{r}^{2}}\dfrac{\partial \rho}{\partial \tau}\; .
	\end{equation}
	Hence, the continuity equation becomes,
	\begin{equation}\label{4.30}
		 \nabla ^{\prime}\rho \cdot \nabla ^{\prime} \tilde{S}+\dfrac{\rho}{2}\nabla ^{\prime 2}\tilde{S} = -m\dfrac{\partial \rho}{\partial \tau}\; .
	\end{equation}

	\subsection{Transformation of the wave functions}

Since the quantities $r,l,p_{r}$ transform under Levi-Civita transformation as given in eqn.(\ref{1141}), eqn.(\ref{1135}) and eqn.(\ref{1136}) respectively, we find
\begin{equation}
	\rho (r,\phi , t) =  \dfrac{C}{\sqrt{rlp_{r}}} 
	=  \dfrac{C}{\sqrt{\tilde{r}^{2}2cl^{\prime}\frac{2c}{\tilde{r}}p_{\tilde{r}}}} = \dfrac{C}{C^{\prime}(2c)} \tilde{\rho}(\tilde{r},\theta , \tau),
\end{equation}
$\tilde{\rho}(\tilde{r},\theta,\tau)$is exactly matching with the one obtained in eqn.(\ref{1160}). Note that $\rho$ and $\tilde{\rho}$ differ only by an irrelevant multiplicative constant, and thus eqn.(\ref{4.30}) becomes,
	\begin{equation}
	\nabla ^{\prime}\tilde{\rho} \cdot \nabla ^{\prime} \tilde{S}+\dfrac{\tilde{\rho}}{2}\nabla ^{\prime 2}\tilde{S} = -m\dfrac{\partial \tilde{\rho}}{\partial \tau},
\end{equation}
which is the continuity equation for the harmonic oscillator.

To transform action, we start from the eqn.(\ref{5.15}) which is,
\begin{equation}
	dS = dR(r) + ld\phi -E dt. \nonumber
\end{equation}
From eqns.(\ref{5.16}) and (\ref{5.17}) it is clear that how $dR(r)$ and $ld\phi$ terms transform under Levi-Civita map. To transform the last term $Edt$, we start from the eqn.(\ref{5.13}) and solve for $\tilde{r}$
\begin{equation}\label{5.33}
	\tilde{r} = \pm \sqrt{\dfrac{4c^{2}E^{\prime}-k}{E+2c^{2}k^{\prime}}}.
\end{equation}
Substituting $\tilde{r}$ into eqn.(\ref{1139}) and simplifying we find,
\begin{equation}
	Edt = 4c(E^{\prime}d\tau)-\left(\dfrac{k}{c}d\tau +2c^{2}k^{\prime}dt\right) = 4c(E^{\prime}d\tau) - \left(\dfrac{k}{c}+2c^{2}k^{\prime}\dfrac{\tilde{r}^{2}}{c}\right)d\tau. \label{5.35}
\end{equation}
 Since the map between the solutions of the 2d Kepler problem and that of the 2d harmonic oscillator is established on constant energy surfaces, here too one expects the mapping of the systems defined on constant energy surfaces. To obtain  the mapping between constant energy surfaces of 2d Kepler and 2d harmonic oscillator, we set the last term on the RHS to be zero and using the relations in eqn.(\ref{500}), we obtain,
\begin{equation}\label{5.36}
	4c^{2}E^{\prime} = E\tilde{r}^{2}.
\end{equation}
Since the total energy of the simple harmonic oscillator is
\begin{equation}\label{5.37}
	E^{\prime} = \dfrac{1}{2}|k^{\prime}|A^{2},
\end{equation}
where $A$ is the amplitude of the oscillation, we get(from eqn.(\ref{500}))
\begin{equation}\label{5.38}
	4c^{2}E^{\prime} = E A^{2}.
\end{equation} 
 the map exists for constant values of $E$ and $E^{\prime}$. This in conjugation with the condition given in eqn.(\ref{5.36}) fixes the value of $\tilde{r}$, further this value should be equal to that of $A$ fixed by eqn.(\ref{5.37}) and eqn.(\ref{5.38}). Hence eqn.(\ref{5.35}) becomes,
\begin{equation}\label{5.39}
	Edt = 4c(E^{\prime}d\tau).
\end{equation}
Thus,
\begin{equation}
	dS = 4c(d\tilde{R}(\tilde{r})+l^{\prime}d\theta-E^{\prime}d\tau) = 4c(d\tilde{S}),
\end{equation} 
which, on integration on both sides, gives
\begin{equation}
	S = 4c \tilde{S}.
\end{equation}
Hence, the wave function for the Kepler problem is transformed to,
\begin{equation}
	\psi = \rho e^{iS} = K \tilde{\rho} (e^{i\tilde{S}})^{4c}
\end{equation}
where RHS is proportional to eqn.(\ref{4.7}) for $c=\frac{1}{4}$. Note that the map between the 2-dimensional Kepler system and
2-dimensional harmonic oscillator is known for $c=\frac{1}{4}$\cite{eh1}.

\subsection{Transformation of Schr\"{o}dinger like equation}

 Schr\"{o}dinger like equation for Kepler problem is
\begin{equation}\label{4.43}
	\left[-\dfrac{1}{2m}\nabla ^{2}-\dfrac{k}{r}+\dfrac{1}{2m}\dfrac{\nabla ^{2}\rho}{\rho}\right]\rho e^{iS} = i \dfrac{\partial }{\partial t}(\rho e^{iS}).
\end{equation}
 Note that $\rho,\tilde{\rho}$ in eqns. (\ref{3.7}) and(\ref{1160}) do not explicitly depend on their respective time parameters, and the time dependence of $S,\tilde{S}$ is given in eqns. (\ref{hpf}) and (\ref{hpf2}), for constant $E$ and $E^{\prime}$ we have,
 \begin{equation}\label{4.44}
 	\dfrac{\partial }{\partial t}(\rho e^{iS}) = -iE(\rho e^{iS}) \; .
 \end{equation}
 Using this in eqn.(\ref{4.43}), we get
 \begin{equation}
 	\left[-\dfrac{1}{2m}\nabla ^{2}-\dfrac{k}{r}+\dfrac{1}{2m}\dfrac{\nabla ^{2}\rho}{\rho}\right]\rho e^{iS} = E(\rho e^{iS}).
 \end{equation}
 Applying the Levi-Civita transformation to the above equation, we get
 \begin{equation}
 	\dfrac{1}{4\tilde{r}^{2}}\left[-\dfrac{1}{2m}\tilde{\nabla} ^{2}-4k+\dfrac{1}{2m}\dfrac{\tilde{\nabla} ^{2}\tilde{\rho}}{\tilde{\rho}}\right]\tilde{\rho}e^{i\tilde{S}}=E\tilde{\rho}e^{i\tilde{S}}\; .
 \end{equation}
 Now using eqn.(\ref{500}) and $c=\dfrac{1}{4}$, the above equation becomes
 \begin{equation}
 	\left[-\dfrac{1}{2m}\tilde{\nabla}^{2}-E^{\prime}+\dfrac{1}{2m}\dfrac{\tilde{\nabla}^{2}\tilde{\rho}}{\tilde{\rho}}\right]\tilde{\rho}e^{i\tilde{S}} = -\dfrac{1}{2}k^{\prime}\tilde{r}^{2}\tilde{\rho}e^{i\tilde{S}} \; .
 \end{equation}
 Rearranging this, we find
  \begin{equation}
 	\left[-\dfrac{1}{2m}\tilde{\nabla}^{2}+\dfrac{1}{2}k^{\prime}\tilde{r}^{2}+\dfrac{1}{2m}\dfrac{\tilde{\nabla}^{2}\tilde{\rho}}{\tilde{\rho}}\right]\tilde{\rho}e^{i\tilde{S}} = E^{\prime}\tilde{\rho}e^{i\tilde{S}} \; ,
 \end{equation}
 which is the Schr\"{o}dinger-like equation for harmonic oscillator potential.

\section{Conclusion}

We have applied the operator formalism of classical mechanics \cite{jc} to two well-studied central potentials: the Kepler problem and the harmonic oscillator in 2-dimensions, and re-derived the well-known equivalence between them.

In the operator formalism \cite{jc} one re-cast systems described by the classical mechanical framework in terms of classical wave function living in the configuration space, operators with well defined action on these wave functions, continuity equation that ensures the particle number conservation and a Schro\"dinger like equation obeyed by the classical wave function. This equation's real and imaginary parts are nothing but the Hamilton-Jacobi equation and the continuity equation satisfied by the classical system. The motivation is to reformulate classical mechanics with the quantum mechanical framework as closely as possible.

In this paper, we start from the Hamilton-Jacobi equation and solve it to obtain Hamilton's principle function, eqns.(\ref{hpf}, \ref{hpf2}). Using this, we set up the continuity equation (see eqn.(\ref{cont})) whose solution gives the probability density associated with the systems under study here (see eqns.(\ref{prob},\ref{1160})). We obtain the classical wave function in eqn.(\ref{wf}) corresponding to Kepler and harmonic potentials in 2-dimensions using this probability density and Hamilton's principle function. We then construct the classical Hamiltonian operator(s) and set up the Schr\"odinger-like equation(s) describing the 2-dimensional Kepler problem and 2-dimensional harmonic oscillator. We show that the normalisation of the classical wave function requires careful treatment, as the wave function is valid only in a finite range. The turning points of the effective potential define this finite range. We have explicitly shown that the classical wave function of the Kepler problem is normalised by imposing the condition
\begin{equation}
\int_{r_-}^{r_+} \psi^*\psi rdrd\phi=1
\end{equation}
where $r_\pm$ is given in eqn.(\ref{rpm}). Similarly, the normalisation condition satisfied by the classical harmonic oscillator's wave function is
\begin{equation}
\int_{{\tilde r}_-}^{{\tilde r}_+} \Phi^*\Phi{\tilde r}d{\tilde r}d\theta=1
\end{equation}
where ${\tilde r}_\pm$ is given in eqn.(\ref{rtildepm}). Notice that the turning points of the 2d Kepler potential $ r_{\pm} = -k/2E \pm k/2E\left(1+2l^2E/(mk^2)\right)$, get mapped to $(\tilde{r}_{\mp})^2$ using eqn.(\ref{1136}) and eqn.(\ref{500}) where $ \tilde{r}_\pm=\left(E^{\prime}/k^{\prime}\pm E^{\prime}/k^{\prime}    \sqrt{1-l^{\prime 2}k^{\prime}/(m E^{\prime 2})}\right)^{1/2}$
is the turning points of the 2d harmonic potential.

We then applied the Levi-Civita map and time to (i) the Hamilton-Jacobi equation describing the 2-dim Kepler problem, in a constant energy surface mapping it to the corresponding equation describing the 2-dim harmonic oscillator, (ii) continuity equation associated with the former and showing that it goes over to that of the later, (iii) classical wave function describing the Kepler problem and mapping it to that of harmonic oscillator, (iv) the Schro\"dinger like equation corresponding to Kepler problem mapping it to that of harmonic oscillator and thereby showing the equivalence of these two models within the operator formalism.

The framework of operator formalism has common features of classical and quantum mechanics, thus turning out to be adept in accommodating the equivalence between the 2-dimensional Kepler problem and the 2-dimensional harmonic oscillator, which is well known in the conventional frameworks of classical and quantum mechanics. Since the operator framework naturally introduces (i) a wave function description of the classical system, and (ii) uncertainty relations akin to those of quantum mechanics between coordinate and momentum operators of the classical model, it is interesting to study the notion of entanglement within this approach. Work along this line is in progress.

\section{Conflict of Interest Statement }

The authors have no conflicts to disclose.

\end{document}